\documentclass[pre,amsmath,amssymb]{revtex4}

\usepackage{wasysym}
\usepackage{graphicx}
\usepackage{dcolumn}
\usepackage{bm}

\begin{document}

\title{Microscopic derivation of self-consistent equations of Anderson localization
 in a disordered medium of finite size}

\author{N. Cherroret}
\email{Nicolas.Cherroret@grenoble.cnrs.fr}
\author{S.E. Skipetrov}
\email{Sergey.Skipetrov@grenoble.cnrs.fr}
\affiliation{Universit\'{e} Joseph Fourier, Laboratoire de Physique et Mod\'{e}lisation des Milieux Condens\'{e}s}
\affiliation{CNRS UMR 5493, B.P. 166, 25 rue des Martyrs,
Maison des Magist\`{e}res, 38042 Grenoble Cedex 09, France}

\date{\today}

\begin{abstract}
We present a microscopic derivation of self-consistent equations of Anderson localization in a disordered medium of finite size. The derivation leads to a renormalized, position-dependent diffusion coefficient. The position dependence of the latter is due to the position dependence of return probability in a bounded medium.  
\end{abstract}

\pacs{42.25.Dd}
\maketitle

\section{Introduction}
The phenomenon of Anderson localization \cite{Anderson} has been studied both experimentally and theoretically for already half a century \cite{thouless74,Ping_Sheng,bart99}. It takes place for waves in strongly disordered media when interference effects become plethoric in the multiple scattering process. Theoretical description of Anderson localization reached a decisive stage in the eighties with the self-consistent (SC) theory of Vollhardt and W\"{o}lfle \cite{Vollhardt}. However, in its original form, this theory did not fully account for finite-size effects. Later, Van Tiggelen {\em et al.} proposed a natural generalization of SC theory to media of finite size by introducing a \emph{position-dependent} diffusion coefficient $D$ \cite{Lagendijk}. This generalized SC theory has been recently used to study the dynamics of Anderson localization in quasi-one-dimensional \cite{SB1} and three-dimensional \cite{SB2} systems. Meanwhile, the generalized SC equations of Refs.\ \onlinecite{Lagendijk,SB1,SB2} have never been derived microscopically. Such a derivation is highly desirable for at least two reasons. First, our recent results indicate that the position dependence of $D$ is crucial for the internal consistency of the theory itself and that some of the important features of Anderson localization (like the $1/L^2$ scaling of the transmission coefficient with the size $L$ of a disordered sample at the mobility edge) cannot be reproduced without fully taking it into account \cite{Nicolas}. Second, the very fact that $D$ should be position-dependent can be questioned in favor of momentum \cite{berk87} or time \cite{berk90} dependencies studied in the past, unless the position dependence of $D$ is given a microscopic justification. This calls for a rigorous derivation of SC equations in a medium of finite size, showing the emergence of position-dependent $D$ from microscopic equations of wave propagation and clarifying the physics behind it. 

In this paper we present a derivation of SC equations of localization in a finite medium
of size $L$ much exceeding the two main ``microscopic'' length scales of the problem: the wavelength
$\lambda$ and the mean free path $\ell$ due to disorder. Our derivation is
based on the ``Hikami box'' formalism \cite{Gorkov,Hikami}. We work in the framework of classical wave scattering, but our results can be extended to quantum particles (e.g., an electron or an atom at low temperatures) described by Schr\"{o}dinger equation with a disordered potential.
Whereas electronic properties of disordered systems have been a subject of intense studies over several decades \cite{Anderson,Vollhardt,berk87,berk90,Gorkov,Hikami,Akkermans_Montambaux,shapiro82,mackinnon94,abrahams79}, the behavior of coherent atomic ensembles (Bose-Einstein condensates) in disordered optical lattices has come into focus only recently \cite{lye05,clement05,shapiro07,skipetrov08}.

Mathematically, the finiteness of the medium comes into play when we evaluate interference corrections to the sum of ladder diagrams. These interference corrections are due to infinite series of maximally-crossed diagrams that we insert inside the ladders. In the presence of time-reversal invariance, the final result depends on the probability for the wave (or quantum particle) to return back to a given point $\mathbf{r}$. Whereas, due to the translational and rotational invariance, this return probability is a position-independent quantity in the infinite medium, it becomes position-dependent in a medium of finite size. In an open medium, the return probability decreases when the boundary of the medium is approached because of the increased probability for the wave to leave the medium through the boundary. This leads to a position-dependent renormalized diffusion coefficient $D$, the renormalization being less important near the boundaries of the disordered medium. The dependence of $D$ on $\mathbf{r}$ is not known in advance, but has to be determined self-consistently by solving a diffusion equation containing the same $D$.  

Anderson localization is often defined as an asymptotic property of eigenstates of disordered wave
(Schr\"{o}dinger, Helmholtz, etc.) equations. The states are said (exponentially) localized if their
intensity decays exponentially at large distances. Another widespread definition of Anderson localization is
vanishing of the diffusion coefficient \cite{bart99}. Strictly speaking, none of these definitions can be directly applied in open media of finite size, which were a subject of extensive work initiated by Thouless \cite{thouless74} and culminated in the scaling theory of localization \cite{abrahams79}. We are not intended to give a review of this work here and the interested reader can find more details in Refs.\ \cite{thouless74,Ping_Sheng,bart99}. For our purposes it will be sufficient to think of ``Anderson localization
in a medium of finite size'' as of an interference, wave phenomenon that would give rise to ``truly localized'' states if the medium were extended to infinity.

The paper is organized as follows. In Sec.\ \ref{I} we review SC theory of localization in infinite and finite media. The main ``building block'' of our derivation --- an ``interference loop'' that we insert inside ladder diagrams to account for interference effects in the intensity Green's function --- is calculated in Sec.\ \ref{II}. In Sec.\ \ref{III} we sum an infinite series of diagrams for the intensity Green's function and obtain the SC equations of localization. Section \ref{IV} is devoted to boundary conditions and a discussion of energy conservation. Finally, we summarize our main results and discuss their implications in Sec.\ \ref{concl}. Technical details of calculations are collected in 4 appendices.    

\section{Theoretical framework}
\label{I}

We consider propagation of a scalar, monochromatic wave of circular frequency $\omega$ in a disordered three-dimensional medium of finite size. The amplitude Green's function $G(\textbf{r},\textbf{r}^{\prime},\omega)$ obeys the Helmholtz equation:
\begin{equation}
[\Delta _{\textbf{r}}+k^{2}(1+\mu(\textbf{r}))]G(\textbf{r},\textbf{r}^{\prime},\omega)=\delta(\textbf{r}-\textbf{r}^{\prime}).\label{Helmholtz}
\end{equation}
Here $\mu(\textbf{r})=\delta\epsilon(\textbf{r})/\bar{\epsilon}$ is the relative fluctuation of the dielectric constant $\epsilon(\textbf{r})=\bar{\epsilon}+\delta\epsilon(\textbf{r})$, $\bar{\epsilon}$ is the average dielectric constant, $k=\sqrt{\bar{\epsilon}}\omega/c$ is the wave number, and $c$ is the speed of wave in a homogeneous medium with $\epsilon = 1$ (vacuum).
We assume that $\mu(\textbf{r})$ obeys the white-noise Gaussian statistics:
 \begin{equation}
k^4\langle\mu(\textbf{r})\mu(\textbf{r}^{\prime})\rangle=\dfrac{4\pi}{\ell}\delta(\textbf{r}-\textbf{r}^{\prime}),
\label{W-NGS}
\end{equation}
where angular brackets denote averaging over realizations of disorder and $\ell$ is the scattering mean free path.
The average amplitude Green's function can be calculated assuming weak disorder ($k \ell \gg 1)$ \cite{Ping_Sheng}:
\begin{equation}
\langle G(\textbf{r},\textbf{r}^{\prime},\omega)\rangle=-\dfrac{1}{4\pi\left| \textbf{r}-\textbf{r}^{\prime}\right|}\exp\left(ik\left|\textbf{r}-\textbf{r}^{\prime}\right|-\dfrac{\left| \textbf{r}-\textbf{r}^{\prime}\right|}{2\ell}\right).
\end{equation}
Although this result has been obtained for the infinite medium, it holds in a medium of finite size as well, provided that the points $\textbf{r}$ and $\textbf{r}^{\prime}$ are at least one mean free path from the boundaries.

In this paper we will be interested in the average intensity Green's function:
\begin{equation}
C(\textbf{r},\textbf{r}^{\prime},\Omega)=\dfrac{4\pi}{c}\langle G(\textbf{r},\textbf{r}^{\prime},\omega_1)G^{*}(\textbf{r},\textbf{r}^{\prime},\omega_2)\rangle,
\label{intensity_GF}
\end{equation}
where $\omega_1=\omega_0+\Omega/2$, $\omega_2=\omega_0-\Omega/2$, and we omit the dependence of $C$ on the carrier frequency $\omega_0$. We assume the latter to be fixed in the remainder of the paper. Physically, the Fourier transform $C(\textbf{r},\textbf{r}^{\prime},t-t^{\prime})$ of Eq.\ (\ref{intensity_GF}) describes the density of wave energy at $\textbf{r}$ at time $t$ due to a short pulse emitted at time $t^{\prime}$ by a point source at $\textbf{r}^{\prime}$. For a quantum particle, $C$ can be interpreted as a probability density of finding the particle in the vicinity of point $\textbf{r}$ at time $t$, provided that the particle was at $\textbf{r}^{\prime}$ at time $t^{\prime}$ (``probability of quantum diffusion'') \cite{Akkermans_Montambaux}.

The analysis of the intensity Green's function is generally complicated and relatively simple results can be obtained only for weak disorder ($k \ell \gg 1$) at large spatial scales ($|\textbf{r} - \textbf{r}^{\prime}| \gg \ell$) and for slow dynamics ($\Omega \ll \omega_0$, $c/\ell$). Under these assumptions, one derives the diffusion equation for the intensity Green's function \cite{Ping_Sheng,Akkermans_Montambaux}:
\begin{equation}
\left( -i\Omega - D_B \Delta_{\textbf{r}}\right)
C(\textbf{r},\textbf{r}^{\prime},\Omega)=\delta(\textbf{r}-\textbf{r}^{\prime}),
\label{eq_for_C_infinite}
\end{equation}
where $D_B=c\ell/3$ is the Boltzmann diffusion coefficient. This equation holds in the infinite as well as in finite media, provided that it is supplemented with appropriate boundary conditions \cite{Ping_Sheng,Akkermans_Montambaux,Zhu} in the latter case. Obviously, Eq.\ (\ref{eq_for_C_infinite}) ignores interference effect and treats the wave as a classical particle that propagates through a disordered medium by diffusion. Vollhardt and W\"{o}lfle \cite{Vollhardt} have shown that interference effects lead to a renormalization of $D_B$ in Eq.\ (\ref{eq_for_C_infinite}). The renormalized diffusion coefficient $D(\Omega)$ obeys \footnote{To lighten the notation, we use $d\textbf{x}$ instead of $d^3\textbf{x}$ to denote three-dimensional integration over a vector $\textbf{x}$.}:
\begin{eqnarray}
	\dfrac{1}{D(\Omega)} &=& \dfrac{1}{D_B}+\dfrac{6\pi}{k^2\ell}
	\int \frac{d\textbf{Q}}{(2\pi)^3} \dfrac{1}{-i\Omega +D(\Omega) \bf{Q}^2}.
\label{eq_for_D_infinite}
\end{eqnarray}
In three dimensions, the integral over $\textbf{Q}$ exhibits an ultraviolet divergence arising from the failure of the diffusion equation (\ref{eq_for_C_infinite}) at small length scales $| \mathbf{r} - \mathbf{r}^{\prime} | < \ell$. This unphysical divergence can be regularized by introducing an upper cutoff of integration $Q_{max} \sim 1/\ell$.

Although, strictly speaking, Eq.\ (\ref{eq_for_C_infinite}) with $D_B$ replaced by $D(\Omega)$ can only be justified for $k \ell \gg 1$, the great success of self-consistent equations (\ref{eq_for_C_infinite}) and (\ref{eq_for_D_infinite}) is due to the fact that they correctly describe many aspects of wave propagation in disordered media all the way down to $k \ell \simeq 1$ (mobility edge) and even at $k \ell < 1$ (Anderson localized regime). In a disordered metal, for example, where the quantity of interest is the dynamic conductivity $\sigma(\Omega) \propto D(\Omega)$, these equations yield the weak localization effect $\sigma(0) \propto 1 - \mathrm{const}/(k\ell)^{2}$, the low-frequency behavior of conductivity at the mobility edge $\sigma(\Omega) \propto (-i \Omega)^{1/3}$ and in the localized (i.e. insulating) phase $\sigma(\Omega) \propto -i \Omega \xi^2$ \cite{shapiro82}. However, not all the results obtained in the framework of SC theory are correct. As an example, we mention the critical exponent $\nu$ describing the divergence of localization length $\xi$ with $k \ell - 1$: $\xi \propto |k\ell - 1|^{-\nu}$. SC theory yields $\nu = 1$, whereas numerical simulations suggest $\nu \approx 1.5$ \cite{mackinnon94}. Another shortcoming of SC theory is its inapplicability to systems with broken time-reversal symmetry.

The derivation of Eq.\ (\ref{eq_for_D_infinite}) heavily relies on the translational invariance and cannot be straightforwardly generalized to media of finite size, even when the size $L$ of the medium is much larger than $\lambda$ and $\ell$. To some extent, Eq.\ (\ref{eq_for_C_infinite}) with $D_B$ replaced by $D(\Omega)$ can still be used to study media of finite size by using a lower cutoff $\sim 1/L$ in the integral over $\mathbf{Q}$ in Eq.\ (\ref{eq_for_D_infinite}) \cite{Vollhardt}. Such an approach can be more or less successful in making qualitative predictions in the spirit of the scaling theory of localization \cite{abrahams79}, but it becomes insufficient when one is interested in fine details of multiple wave scattering close to the mobility edge and in the localized regime: coherent backscattering cone \cite{Lagendijk}, dynamics of short pulses \cite{SB1,SB2}, or precise scaling of the transmission coefficient with the size of disordered sample \cite{Nicolas}. A plausible generalization of SC theory to media of finite size can be obtained by noticing that, by virtue of Eq.\ (\ref{eq_for_C_infinite}), the $\mathbf{Q}$-integral of Eq.\ (\ref{eq_for_D_infinite}) is formally equal to the ``return probability'' $C(\mathbf{r}, \mathbf{r}, \Omega)$. We can therefore rewrite Eq.\ (\ref{eq_for_D_infinite}) as $1/D(\Omega) =  1/D_B + 6\pi/(k^2\ell) C(\textbf{r},\textbf{r},\Omega)$. Van Tiggelen \emph{et al.} conjectured \cite{Lagendijk} that in this new form the self-consistent equation for $D$ might hold in a medium of finite size as well. In a medium of finite size, the position dependence of $C(\mathbf{r}, \mathbf{r}, \Omega)$ naturally gives rise to a position dependence of $D$:
\begin{eqnarray}
	\dfrac{1}{D(\mathbf{r},\Omega)} &=&  \dfrac{1}{D_B}+\dfrac{6\pi}{k^2\ell}C(\textbf{r},\textbf{r},\Omega).
\label{eq_for_D_finite}
\end{eqnarray}
If we then enforce diffusive behavior of the intensity Green's function and insist on the energy conservation, the equation for $C$ becomes \cite{Lagendijk}
\begin{equation}
\left( -i\Omega -\boldsymbol{\nabla}_{\textbf{r}} \cdot D(\textbf{r},\Omega) \boldsymbol{\nabla}_{\textbf{r}}\right)
C(\textbf{r},\textbf{r}^{\prime},\Omega)=\delta(\textbf{r}-\textbf{r}^{\prime}).
\label{eq_for_C_finite}
\end{equation}

Although SC equations (\ref{eq_for_D_finite}) and (\ref{eq_for_C_finite}) appear to be a powerful tool to study Anderson localization in realistic situations \cite{Lagendijk,SB1,SB2,Nicolas}, they still remain a conjecture and lack microscopic justification. Derivation of these equations from the first principles is the main purpose of the present paper.

\section{Interference effects in finite media}
\label{II}
Formally, the intensity Green's function is given by \cite{Lagendijk_WL}
\begin{eqnarray}
C(\textbf{r},\textbf{r}^{\prime},\Omega) &=& \dfrac{4\pi}{c}\langle G(\textbf{r},\textbf{r}^{\prime},\omega_1)\rangle \langle G^*(\textbf{r},\textbf{r}^{\prime},\omega_2)\rangle \nonumber \\
&+& \dfrac{4\pi}{c}\int d\textbf{r}_1d\textbf{r}_2d\textbf{r}_3d\textbf{r}_4\langle G(\textbf{r},\textbf{r}_1,\omega_1)\rangle \langle G^*(\textbf{r},\textbf{r}_3,\omega_2)\rangle
\nonumber \\
&\times& \Gamma(\textbf{r}_1,\textbf{r}_2,\textbf{r}_3,\textbf{r}_4,\Omega)\langle G(\textbf{r}_2,\textbf{r}^{\prime},\omega_1)\rangle \langle G^*(\textbf{r}_4,\textbf{r}^{\prime},\omega_2)\rangle, \label{Rigorous_intensity}
\end{eqnarray}
where $\Gamma (\textbf{r}_1,\textbf{r}_2,\textbf{r}_3,\textbf{r}_4,\Omega)$ is the complete vertex function given by a sum of all diagrams connecting scattering paths corresponding to $G$ and $G^*$. The first term $\langle G\rangle\langle G^*\rangle$ on the right-hand side (r.h.s.) of Eq.\ (\ref{Rigorous_intensity}) will be neglected in the following. Indeed, it is exponentially small at large distances $\vert\textbf{r}-\textbf{r}^{\prime}\vert \gg \ell$ that are of main interest for us here.

\begin{figure}[t]
\includegraphics[width=12.0cm]{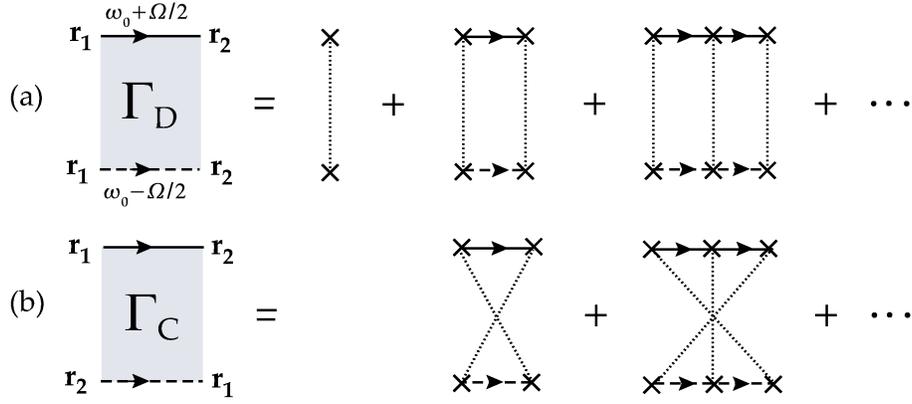}
\caption{\label{Ladder_diagrams} (a) Sum of ladder diagrams $\Gamma_D(\textbf{r}_1,\textbf{r}_2,\Omega)$ and (b) sum of maximally-crossed diagrams $\Gamma_C(\textbf{r}_1,\textbf{r}_2,\Omega)$.  Solid and dashed lines denote $\langle G\rangle$ and $\langle G^*\rangle$, respectively. The dotted line symbolizes the correlation function of disorder
$k^4 \langle \mu(\mathbf{r}) \mu(\mathbf{r}^{\prime}) \rangle$ given by Eq. (\ref{W-NGS}). Crosses denote scattering events. Integrations over positions of all internal scattering events are assumed. In all diagrams of this paper, $\langle G\rangle$ and $\langle G^*\rangle$ should be evaluated at frequencies $\omega_1 = \omega_0 + \Omega/2$ and  $\omega_2 = \omega_0 - \Omega/2$, respectively. We show this explicitly in the panel (a) of this figure only.}  
\end{figure}

In the regime of weak disorder, defined by $k\ell\gg 1$, $\Gamma (\textbf{r}_1,\textbf{r}_2,\textbf{r}_3,\textbf{r}_4,\Omega)=\delta(\textbf{r}_1-\textbf{r}_3)\delta(\textbf{r}_2-\textbf{r}_4)\Gamma_D (\textbf{r}_1,\textbf{r}_2,\Omega)$ with $\Gamma_D$ a sum of ladder diagrams \cite{Ping_Sheng,Akkermans_Montambaux,Vollhardt} shown in Fig. \ref{Ladder_diagrams}(a). We denote $C$ given by Eq.\ (\ref{Rigorous_intensity}) with $\Gamma_D$ substituted for $\Gamma$ by $C_D$. At large distances $| \mathbf{r} - \mathbf{r}^{\prime}| \gg \ell$ and in the limit of small $\Omega$, $C_D$ obeys the diffusion equation (\ref{eq_for_C_infinite}).
We also  introduce a sum of maximally-crossed diagrams $\Gamma_C(\textbf{r}_1,\textbf{r}_2,\Omega)$ shown in Fig.\ \ref{Ladder_diagrams}(b). If we do not consider the first term on the r.h.s. of Fig.\ \ref{Ladder_diagrams}(a), we can formally obtain $\Gamma_C$  from $\Gamma_D$ by rotating the bottom propagation line of the diagram of Fig.\ \ref{Ladder_diagrams}(a) by 180$^{\circ}$ in the plane perpendicular to the plane of the figure.  The time-reversal invariance, that we assume to hold throughout this paper, implies $\Gamma_C(\textbf{r}_1,\textbf{r}_2,\Omega) = \Gamma_D (\textbf{r}_1,\textbf{r}_2,\Omega)$ if $|\mathbf{r}_1 - \mathbf{r}_2|$ exceeds the correlation length of disorder (i.e. if $\mathbf{r}_1 \ne \mathbf{r}_2$ for the white-noise disorder that we consider here) because the first term of Fig.\ \ref{Ladder_diagrams}(a) can be neglected in this case.

\begin{figure}[t]
\includegraphics[width=6.0cm]{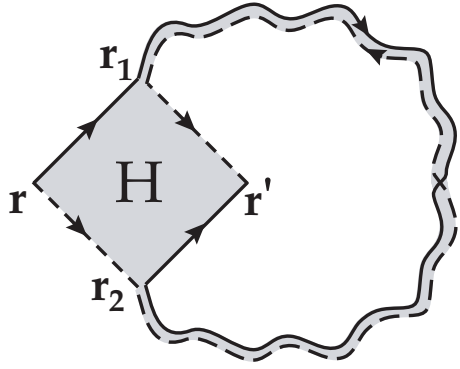}
\caption{\label{Hikami_boxes} The diagram $X(\textbf{r},\textbf{r}^{\prime},\Omega)$ that we use to introduce interference effects in the calculation of intensity Green's function. This diagram is made of a four-point Hikami box $H(\textbf{r},\textbf{r}_1,\textbf{r}^{\prime},\textbf{r}_2)$ --- detailed in Appendix \ref{A} --- and of the sum of maximally-crossed diagrams $\Gamma_C(\textbf{r}_1,\textbf{r}_2,\Omega)$ shown by wavy lines connecting $\textbf{r}_1$ and $\textbf{r}_2$.}  
\end{figure}

To account for interference effects during propagation, we consider a loop-shaped diagram $X(\textbf{r},\textbf{r}^{\prime},\Omega)$ shown in Fig.\ \ref{Hikami_boxes}. This diagram is made of a square diagram known as a four-point Hikami box $H(\textbf{r},\textbf{r}_1,\textbf{r}^{\prime},\textbf{r}_2)$ \cite{Gorkov, Hikami} and of a sum of maximally-crossed diagrams $\Gamma_C(\textbf{r}_1,\textbf{r}_2,\Omega)$ that we replace by $\Gamma_D(\textbf{r}_1,\textbf{r}_2,\Omega)$, making use of time-reversal invariance:
\begin{equation}
X(\textbf{r},\textbf{r}^{\prime},\Omega) = \int d\textbf{r}_1d\textbf{r}_2H(\textbf{r},\textbf{r}_1,\textbf{r}^{\prime},\textbf{r}_2)\Gamma_D(\textbf{r}_1,\textbf{r}_2,\Omega).
\label{calculation_loop}
\end{equation}
Because $H$ is a local object having non-zero value only when all the 4 points $\textbf{r}$, $\textbf{r}_1$, $\textbf{r}^{\prime}$ and $\textbf{r}_2$ are within a distance of order $\ell$ from each other, we can expand $\Gamma_D$ in series around $\mathbf{r}$, assuming that its spatial variations are small at the scale of $\ell$:
\begin{eqnarray}
\Gamma_D(\textbf{r}_1,\textbf{r}_2,\Omega) &\simeq & \Big\{1+(\textbf{r}_1-\textbf{r})\cdot\boldsymbol{\nabla}_{\textbf{r}_1} 
+(\textbf{r}_2-\textbf{r})\cdot\boldsymbol{\nabla}_{\textbf{r}_2}
\nonumber\\
&+& \dfrac{1}{2}{\left[(\textbf{r}_1-\textbf{r})\cdot\boldsymbol{\nabla}_{\textbf{r}_1} \right]}^2 
+\dfrac{1}{2}{\left[(\textbf{r}_2-\textbf{r})\cdot\boldsymbol{\nabla}_{\textbf{r}_2} \right]}^2 + \dots \Big\} \Gamma_D(\textbf{r}_1,\textbf{r}_2,\Omega)\mid _{\textbf{r}_1=\textbf{r}_2=\textbf{r}}.
\label{expansion_gamma}
\end{eqnarray}
We will truncate this expansion to the first order and use the reciprocity principle $\Gamma_D(\textbf{r}_1,\textbf{r}_2,\Omega)=\Gamma_D(\textbf{r}_2,\textbf{r}_1,\Omega)$ that allows us to rewrite Eq. (\ref{expansion_gamma}) as
\begin{equation}
\Gamma_D(\textbf{r}_1,\textbf{r}_2,\Omega)\simeq \left[ 1+\dfrac{1}{2}(\textbf{r}_1+\textbf{r}_2-2\textbf{r})\cdot\boldsymbol{\nabla}_{\textbf{r}} \right]
\Gamma_D(\textbf{r},\textbf{r},\Omega).
\label{expansion_simplified}
\end{equation}
Substituting this into Eq. (\ref{calculation_loop}) we obtain
\begin{equation}
X(\textbf{r},\textbf{r}^{\prime},\Omega)=
\left[ H(\textbf{r},\textbf{r}^{\prime})
+\dfrac{1}{2} \textbf{H}_f(\textbf{r},\textbf{r}^{\prime})\cdot\boldsymbol{\nabla}_{\textbf{r}} \right] \Gamma_D(\textbf{r},\textbf{r},\Omega)
\label{Cloop},
\end{equation}
with
\begin{equation}
H(\textbf{r},\textbf{r}^{\prime})=\int d\textbf{r}_1d\textbf{r}_2H(\textbf{r},\textbf{r}_1,\textbf{r}^{\prime},\textbf{r}_2)
\label{H_definion}
\end{equation}
and
\begin{equation}
\textbf{H}_f(\textbf{r},\textbf{r}^{\prime})=\int d\textbf{r}_1d\textbf{r}_2(\textbf{r}_1+\textbf{r}_2-2\textbf{r})H(\textbf{r},\textbf{r}_1,\textbf{r}^{\prime},\textbf{r}_2).
\label{H_f_definition}
\end{equation}
The first term on the r.h.s. of Eq.\ (\ref{Cloop}) is the ``usual'' term arising in the infinite medium as well \cite{Akkermans_Montambaux}. The second term on the r.h.s. is non-zero only in a finite medium because $\Gamma_D(\textbf{r},\textbf{r},\Omega)$ is independent of $\mathbf{r}$ in the infinite medium. It will be seen from the following that this term is of fundamental importance for the derivation of self-consistent equations of localization in a finite medium.

A calculation detailed in Appendix \ref{A} gives
\begin{equation}
\textbf{H}_f(\textbf{r},\textbf{r}^{\prime})=-(\textbf{r}-\textbf{r}^{\prime})H(\textbf{r},\textbf{r}^{\prime}).
\label{H_f_result}
\end{equation}
Substituting Eq. (\ref{H_f_result}) into Eq. (\ref{Cloop}) we obtain
\begin{equation}
X(\textbf{r},\textbf{r}^{\prime},\Omega)=H(\textbf{r},\textbf{r}^{\prime})\left[1 - \dfrac{1}{2}(\textbf{r}-\textbf{r}^{\prime}) \cdot \boldsymbol{\nabla}_{\textbf{r}} \right] \Gamma_D(\textbf{r},\textbf{r},\Omega).
\label{Cloop_r_r'}
\end{equation}

For convenience of calculations, we introduce the difference variable $\Delta\textbf{r}=\textbf{r}-\textbf{r}^{\prime}$, such that a given function $f$ of $\textbf{r}$ and $\textbf{r}^{\prime}$ becomes a function $\tilde{f}$ of $\textbf{r}$ and $\Delta\textbf{r}$. In particular, $H(\textbf{r},\textbf{r}^{\prime})$ becomes $\tilde{H}(\Delta \mathbf{r})$ and does not depend on $\textbf{r}$ \cite{Akkermans_Montambaux}. Using the new set of variables $\textbf{r}$ and $\Delta\textbf{r}$, we have $\Gamma_D(\textbf{r},\textbf{r},\Omega)=\tilde{\Gamma}_D(\textbf{r},\Delta\textbf{r}=\textbf{0},\Omega)$. Equation (\ref{Cloop_r_r'}) becomes
\begin{equation}
\tilde{X}(\textbf{r},\Delta\textbf{r},\Omega) = \tilde{H}(\Delta\textbf{r})\left[1-\dfrac{1}{2}\Delta\textbf{r} \cdot \boldsymbol{\nabla}_{\textbf{r}}\right]
\tilde{\Gamma}_D(\textbf{r},\textbf{0},\Omega).
\label{Cloop_R_Dr}
\end{equation}
We now take the Fourier transform of Eq. (\ref{Cloop_R_Dr}) with respect to $\Delta\textbf{r}$ and consider the limit $\textbf{q} \rightarrow \textbf{0}$. Because the Fourier transform $\tilde{H}(\textbf{q})$ of $\tilde{H}(\Delta\textbf{r})$ is equal to $D_B\ell^4q^2/8\pi c k^2$ in this limit \cite{Akkermans_Montambaux, Feng}, we obtain
\begin{equation}
\tilde{X}(\textbf{r},\textbf{q},\Omega)=\dfrac{-\ell^4D_B}{8 \pi c k^2}\left[(i\textbf{q})^2+(i\textbf{q}) \cdot \boldsymbol{\nabla}_{\textbf{r}}\right]
\tilde{\Gamma}_D(\textbf{r},\textbf{0},\Omega).
\label{Cloop_r_Dr_Fourier}
\end{equation}
An approximate expression for $\tilde{X}(\textbf{r},\Delta\textbf{r},\Omega)$ can then be obtained by the inverse Fourier transform of Eq. (\ref{Cloop_r_Dr_Fourier}) with respect to $\textbf{q}$ (see Appendix \ref{B}):
\begin{equation}
\tilde{X}(\textbf{r},\Delta\textbf{r},\Omega)=\dfrac{-\ell^4D_B}{8 \pi c k^2}
\left\{
\left[ \Delta_{\Delta\mathbf{r}} \delta(\Delta \mathbf{r}) \right] +
\left[ \boldsymbol{\nabla}_{\Delta\mathbf{r}} \delta(\Delta \mathbf{r}) \right]
\cdot \boldsymbol{\nabla}_{\mathbf{r}}
\right\} 
\tilde{\Gamma}_D(\textbf{r},\textbf{0},\Omega).
\label{Cloop_r_Dr_explicit}
\end{equation}
Because $\boldsymbol{\nabla}_{\Delta\mathbf{r}} \delta(\Delta \mathbf{r}) =
\boldsymbol{\nabla}_{\mathbf{r}} \delta(\mathbf{r} - \mathbf{r}^{\prime})$ and
$\Delta_{\Delta\mathbf{r}} \delta(\Delta \mathbf{r}) =
\Delta_{\mathbf{r}} \delta(\mathbf{r} - \mathbf{r}^{\prime})$,
Eq.\ (\ref{Cloop_r_Dr_explicit}) can be rewritten in terms of the original variables $\textbf{r}$ and $\textbf{r}^{\prime}$ as
\begin{equation}
X(\textbf{r},\textbf{r}^{\prime},\Omega)=\dfrac{-\ell^4D_B}{8 \pi c k^2}\boldsymbol{\nabla}_{\textbf{r}} \cdot \left[\Gamma_D(\textbf{r},\textbf{r},\Omega)\boldsymbol{\nabla}_{\textbf{r}}\right]\delta(\textbf{r}-\textbf{r}^{\prime}).
\label{Cloop_r_r'_explicit}
\end{equation}

\section{Derivation of self-consistent equations}
\label{III}

\begin{figure}[t]
\includegraphics[width=16cm]{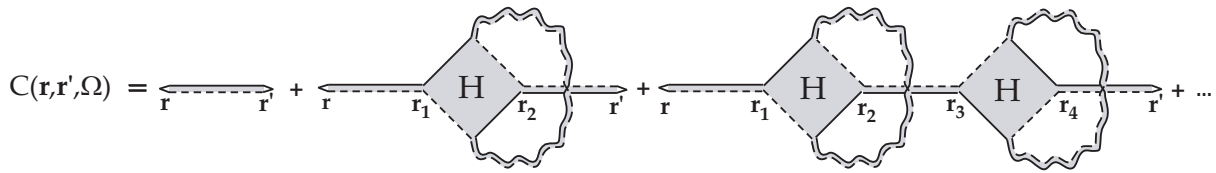}
\caption{\label{Series_diagrams}
Diagrammatic representation of an infinite series of diagrams contributing to the intensity Green's function. The first term is the sum of ladder diagrams. The second term is the sum of ladder diagrams with a single interference loop denoted by wavy lines and equal to an infinite sum of maximally-crossed diagrams. Next terms contain 2, 3, etc. consecutive interference loops. The ladder and the maximally-crossed diagrams are joined together by a Hikami box detailed in Appendix \ref{A}. The analytic representation of this diagrammatic series is given by Eq.\ (\ref{Resummation}).}  
\end{figure}

We will now use the diagram $X$ of Fig.\ \ref{Hikami_boxes} analyzed in the previous section to include interference effects in the calculation of intensity Green's function $C(\textbf{r},\textbf{r}^{\prime},\Omega)$. To this end, we insert the ``interference loop'' $X$ in the sum of ladder diagrams for $C_D$ and account for the possibility of having multiple consecutive interference loops. This leads to an infinite series of diagrams shown in Fig.\ \ref{Series_diagrams}. This series can be written analytically as
\begin{eqnarray}
C(\textbf{r},\textbf{r}^{\prime},\Omega) &=& C_D(\textbf{r},\textbf{r}^{\prime},\Omega) + 
\frac{4 \pi c}{\ell^2} \int C_D(\textbf{r},\textbf{r}_1,\Omega)X(\textbf{r}_1,\textbf{r}_2,\Omega)C_D(\textbf{r}_2,\textbf{r}^{\prime},\Omega)d\textbf{r}_1 d\textbf{r}_2\nonumber\\
&+& \left( \frac{4 \pi c}{\ell^2} \right)^2 \int C_D(\textbf{r},\textbf{r}_1,\Omega)X(\textbf{r}_1,\textbf{r}_2,\Omega)C_D(\textbf{r}_2,\textbf{r}_3,\Omega)
\nonumber \\
&\times& X(\textbf{r}_3,\textbf{r}_4,\Omega)C_D(\textbf{r}_4,\textbf{r}^{\prime},\Omega)d\textbf{r}_1 d\textbf{r}_2d\textbf{r}_3d\textbf{r}_4
+ \dots
\label{Resummation}
\end{eqnarray}
 
We now apply the operator $-i\Omega-D_B \Delta_{\textbf{r}}$ to Eq. (\ref{Resummation}) and use Eq.\ (\ref{eq_for_C_infinite}) for $C_D$ and Eq. (\ref{Cloop_r_r'_explicit}) for $X(\textbf{r},\textbf{r}^{\prime})$. This yields (see the detailed calculation in Appendix \ref{C}):
\begin{equation}
\left[-i\Omega-D_B \Delta_{\textbf{r}} \right]C(\textbf{r},\textbf{r}^{\prime},\Omega)=\delta(\textbf{r}-\textbf{r}^{\prime})-\dfrac{\ell^2D_B}{2k^2}\boldsymbol{\nabla}_{\textbf{r}}\cdot\left[\Gamma_D(\textbf{r},\textbf{r},\Omega)\boldsymbol{\nabla}_{\textbf{r}}C(\textbf{r},\textbf{r}^{\prime},\Omega) \right],
\label{applied_operator_init}
\end{equation}
or
\begin{equation}
\left[-i\Omega-\boldsymbol{\nabla}_{\textbf{r}}\cdot\left(D_B-\dfrac{\ell^2D_B}{2k^2}\Gamma_D(\textbf{r},\textbf{r},\Omega)\right)\boldsymbol{\nabla}_{\textbf{r}}\right]C(\textbf{r},\textbf{r}^{\prime},\Omega)=\delta(\textbf{r}-\textbf{r}^{\prime}).
\label{applied_operator}
\end{equation}
As we demonstrate in Appendix \ref{D}, $\Gamma_D$ is proportional to $C_D$:
$\Gamma_D(\textbf{r},\textbf{r},\Omega)=(4\pi c/\ell^2) C_D({\textbf{r},\textbf{r}},\Omega)$.
This allows us to define a renormalized, position-dependent diffusion coefficient 
\begin{equation}
D(\textbf{r},\Omega)=D_B-\dfrac{2\pi c}{k^2}D_B C_D(\textbf{r},\textbf{r},\Omega).
\label{diffusion_coeff_WL}
\end{equation}
and rewrite Eq.\ (\ref{applied_operator}) as
\begin{equation}
\left[-i\Omega-\boldsymbol{\nabla}_{\textbf{r}} \cdot D(\textbf{r},\Omega)\boldsymbol{\nabla}_{\textbf{r}}\right]C(\textbf{r},\textbf{r}^{\prime},\Omega)=\delta(\textbf{r}-\textbf{r}^{\prime}).
\label{equation_for_C}
\end{equation}

The last step consists in applying the self-consistency principle \cite{Vollhardt}. This can be done by using $D(\textbf{r},\Omega)$ instead of $D_B$ when calculating the second term on the r.h.s. of Eq. (\ref{diffusion_coeff_WL}). Diagrammatically, this procedure is equivalent to inserting ``secondary loops'' in the loops shown by wavy lines in Fig.\ \ref{Series_diagrams} and then inserting the same loops in these secondary loops, etc., thus obtaining a sum of diagrams with an infinite sequence of loops inserted one inside the other. Physically, this simply means that the same, self-consistent diffusion coefficient $D(\mathbf{r}, \Omega)$ should be used when we calculate the intensity Green's function $C$ and the sum of maximally-crossed diagrams $\Gamma_C$. More specifically, we have to perform the following replacements:
\begin{enumerate}
\item We replace $D_B$ by $D$ in $H(\textbf{r},\textbf{r}^{\prime})$ in Eq. (\ref{Cloop_r_r'}), or equivalently in $H(\textbf{q})$, such that $D_B$ is replaced by $D$ in the second term on the r.h.s. of Eq. (\ref{diffusion_coeff_WL}).
\item We replace $D_B$ by $D$ in $\Gamma_D$ in Eq. (\ref{Cloop_r_r'}), which amounts to replace $C_D$ by $C$ in the second term on the r.h.s. of Eq. (\ref{diffusion_coeff_WL}).
\end{enumerate}
Equation (\ref{diffusion_coeff_WL}) then becomes
$D(\textbf{r},\Omega)=D_B-(2\pi c/k^2) D(\mathbf{r},\Omega) C(\textbf{r},\textbf{r},\Omega)$ or
\begin{equation}
\dfrac{1}{D(\textbf{r},\Omega)}=\dfrac{1}{D_B}+\dfrac{6\pi}{k^2\ell}C(\textbf{r},\textbf{r},\Omega).
\label{diffusion_coeff}
\end{equation}
This completes the derivation of self-consistent equations of localization --- Eqs.\ (\ref{equation_for_C}) and (\ref{diffusion_coeff}) --- in a medium of finite size.

	The solution of the diffusion equation (\ref{equation_for_C}) in three dimensions diverges when $\mathbf{r}^{\prime} \rightarrow \mathbf{r}$: $C(\textbf{r},\textbf{r}^{\prime},\Omega) \propto 1/|\mathbf{r} - \mathbf{r}^{\prime}|$. This unphysical divergence poses potential problems in Eq.\ (\ref{diffusion_coeff}) that contains $C(\textbf{r},\textbf{r},\Omega)$. One possibility to regularize this divergence is to represent $C(\textbf{r},\textbf{r}^{\prime},\Omega)$ as a Fourier transform of $C(\textbf{r},\textbf{q},\Omega)$, where $\mathbf{q}$ is a variable conjugated to $\Delta \mathbf{r} = \mathbf{r} - \mathbf{r}^{\prime}$, and then cut off the integration over $\mathbf{q}$ at some $q_{max} \sim 1/\ell$. The exact proportionality constant between $q_{max}$ and $1/\ell$ will determine the exact position of the mobility edge $k \ell \sim 1$. It is also possible to cut off only the integration over $\mathbf{q}_{\perp} = (q_x, q_y)$, leaving the integration over $q_z$ unrestricted. Such a two-dimensional cutoff is easier to implement for the particular geometry of a disordered slab perpendicular to the $z$ axis \cite{SB2,Nicolas}. As could be expected, the main qualitative features of final results are largely insensitive to the details of the large-$q$ cutoff, although quantitative details can vary slightly. 

\section{Energy conservation and boundary conditions}
\label{IV}

It is important to note that although we have obtained Eq.\ (\ref{equation_for_C}) by summing only the diagrams of certain type and neglecting many other diagrams, this equation satisfies the conservation of energy \emph{exactly}. Indeed, let us take its inverse Fourier transform with respect to $\Omega$:
\begin{equation}
\dfrac{\partial C(\textbf{r},\textbf{r}^{\prime},t)}{\partial t}-\int \frac{d\Omega}{2 \pi}\boldsymbol{\nabla}_{\textbf{r}} \cdot D(\textbf{r},\Omega)\boldsymbol{\nabla}_{\textbf{r}}C(\textbf{r},\textbf{r}^{\prime},\Omega) e^{-i\Omega t}=\delta(\textbf{r}-\textbf{r}^{\prime})\delta(t).
\label{continuity_equation}
\end{equation}
The flux of energy is given by Fick's law: $\textbf{J}(\textbf{r},\textbf{r}^{\prime},t) = -\int d\Omega/(2\pi) D(\textbf{r},\Omega)\boldsymbol{\nabla}_{\textbf{r}}C(\textbf{r},\textbf{r}^{\prime},\Omega)e^{-i\Omega t}$. By integrating Eq.\ (\ref{continuity_equation}) over a control volume $V$ contained inside the disordered medium and enclosed by a surface $S$, we obtain
\begin{equation}
\int_V \dfrac{\partial C(\textbf{r},\textbf{r}^{\prime},t)}{\partial t}d\mathbf{r}=-\int_V\boldsymbol{\nabla}_{\textbf{r}}
 \cdot \textbf{J}(\textbf{r},\textbf{r}^{\prime},t)d\mathbf{r}+\delta(t) \int_V \delta(\mathbf{r}-\mathbf{r}^{\prime})d\mathbf{r}.
\label{Ostrogradsky}
\end{equation}
We now apply the Gauss-Ostrogradsky theorem to the first term on the r.h.s. of Eq. (\ref{Ostrogradsky}) and assume that the source point  $\mathbf{r}^{\prime}$ is contained inside $V$:
\begin{equation}
\dfrac{d}{dt}\int_V C(\textbf{r},\textbf{r}^{\prime},t)d\mathbf{r}=-\oint_S\textbf{J}(\textbf{r},\textbf{r}^{\prime},t) \cdot d\textbf{S}+\delta(t).
\label{energy_conservation}
\end{equation}
Here $d\mathbf{S}$ is a vector normal to the surface element $dS$ and directed outwards the volume $V$.

Equation (\ref{energy_conservation}) is a conservation equation. It states that the variation of wave energy in the volume $V$ is given by a balance of energy emitted by the source (the second term on the r.h.s.) and energy leaving the volume through its surface $S$ (the first term on the r.h.s.).

Although inside a disordered medium the energy flux $\mathbf{J}(\mathbf{r}, \mathbf{r}^{\prime},t)$ can have arbitrary magnitude and direction consistent with the diffusion equation (\ref{equation_for_C}) and Fick's law, additional factors come into play at the surface of the medium. More specifically, for an open disordered medium of convex shape surrounded by the free space, no energy flux enters the medium from outside, provided that all sources are located inside the medium. This simple principle allows a derivation of boundary conditions for the intensity Green's function at the surface of disordered medium.
Following Zhu \emph{et al.} \cite{Zhu}, we consider a disordered medium occupying the half-space $z > 0$.
 At a given point $\textbf{r}$ inside the medium, the Fourier component of intensity $I(\textbf{u},\textbf{r},\textbf{r}^{\prime},\Omega)$ propagating in the direction of a unit vector $\textbf{u}$, can be represented as \cite{Akkermans_Montambaux,Zhu}
\begin{eqnarray}
I(\textbf{u},\textbf{r},\textbf{r}^{\prime},\Omega) &=& C(\textbf{r},\textbf{r}^{\prime},\Omega)+\dfrac{3}{c}\textbf{J}(\textbf{r},\textbf{r}^{\prime},\Omega)\cdot\textbf{u}
\nonumber \\
&=& C(\textbf{r},\textbf{r}^{\prime},\Omega)-\dfrac{3}{c}D(\textbf{r},\Omega)\boldsymbol{\nabla}_\textbf{r}C(\textbf{r},\textbf{r}^{\prime},\Omega)\cdot\textbf{u},
\label{I(u,r)}
\end{eqnarray}
where Fick's law was used to obtain the second line.
The total flux of wave energy crossing some plane $z=\textrm{const}$ at point $\mathbf{r}$ in the positive direction of axis $z$ is
\begin{equation}
J_+(\textbf{r},\textbf{r}^{\prime},\Omega)=\dfrac{c}{4\pi}\int_0^{2\pi}d\phi \int_0^{\pi/2}d\theta \sin\theta\; u_z I(\textbf{u},\textbf{r},\textbf{r}^{\prime},\Omega),
\label{J(r)}
\end{equation}
where $u_z=\cos\theta$ is the $z$ component of $\textbf{u}$. We then substitute Eq. (\ref{I(u,r)}) into Eq. (\ref{J(r)}) and perform integrations over $\theta$ and $\phi$. This yields 
\begin{equation}
J_+(\textbf{r},\textbf{r}^{\prime},\Omega)=\dfrac{C(\textbf{r},\textbf{r}^{\prime},\Omega) c}{4}-\dfrac{D(\textbf{r},\Omega)}{2} \dfrac{\partial C(\textbf{r},\textbf{r}^{\prime},\Omega)}{\partial z}.
\end{equation}
By requiring $J_+(\textbf{r},\textbf{r}^{\prime},\Omega)=0$ at the surface $z=0$ of the medium, we obtain the following boundary condition:
\begin{equation}
\left. C(\textbf{r},\textbf{r}^{\prime},\Omega)\right|_{z=0}-\dfrac{2}{c}\left.D(\textbf{r},\Omega)\right|_{z=0}\left. \dfrac{\partial C(\textbf{r},\textbf{r}^{\prime},\Omega)}{\partial z}\right|_{z=0}=0.
\label{boundary_condition}
\end{equation} 

For a medium of more complex but still convex shape, the above derivation can be repeated locally in the vicinity of each point of the medium surface $S$, assumed to be locally flat. This yields
\begin{equation}
C(\textbf{r},\textbf{r}^{\prime},\Omega)-\dfrac{2}{3} \ell \frac{D(\textbf{r},\Omega)}{D_B} \left(\mathbf{n}(\mathbf{r}) \cdot \boldsymbol{\nabla} \right) C(\textbf{r},\textbf{r}^{\prime},\Omega)=0,
\label{boundary_condition2}
\end{equation} 
where $\mathbf{n}(\mathbf{r})$ is a unit inward normal to the surface $S$ at the point $\mathbf{r} \in S$. This equation is the boundary condition for the intensity Green's function at an open boundary. It can be generalized to include internal reflections of waves at the boundary by replacing $2\ell/3$ by a larger ``extrapolation length'' $z_0$ in front of the second term on its l.h.s., in complete analogy with Ref.\ \onlinecite{Zhu}. 

\section{Conclusion}
\label{concl}

In this paper we derived the self-consistent (SC) equations of Anderson localization --- Eqs.\ (\ref{equation_for_C}) and (\ref{diffusion_coeff}) --- starting from the first principles. Mathematically, this was achieved by dressing the ladder propagator with ``interference loops'' made of maximally-crossed diagrams. Each loop was inserted into the ladder with the help of a Hikami-box diagram. The SC equations were then obtained by applying the self-consistency principle.

The essential difference of our derivation compared to the derivation of SC equations in the infinite medium is the position dependence of the sum of ladder diagrams $\Gamma_D(\mathbf{r}, \mathbf{r}^{\prime}, \Omega)$ with coinciding end points $\mathbf{r} = \mathbf{r}^{\prime}$. This position dependence leads to the appearance of an additional term, proportional to $\boldsymbol{\nabla}_{\mathbf{r}} \Gamma_D(\mathbf{r}, \mathbf{r}, \Omega)$, in a series expansion of  $\Gamma_D(\mathbf{r}_1, \mathbf{r}_2, \Omega)$ around an arbitrary point $\mathbf{r}$. As a consequence, we have to keep an additional term in the expression of Hikami box employed to connect ladder and maximally-crossed diagrams in our approach. It is this term that finally allows us to derive SC equations of localization in a medium of finite size.

Although the condition $k \ell \gg 1$ was explicitly used to derive Eqs.\ (\ref{equation_for_C}) and (\ref{diffusion_coeff}), one can still hope that, similarly to SC equations in the infinite medium, they could yield reasonable results in the vicinity of mobility edge and in the localized regime. According to Refs.\ \onlinecite{Lagendijk,SB1,SB2,Nicolas}, this seems indeed be the case. However, one should understand that even though the general form of these equations might be largely universal in both diffuse and localized regimes, the numerical prefactor $6\pi/k^2 \ell$ in front of the second term in the SC equation for $D(\mathbf{r}, \Omega)$, Eq.\ (\ref{diffusion_coeff}), should not be taken too seriously because it originates from the calculation of complicated diagrams that was carried out in the limit $k \ell \gg 1$ only (see Appendix \ref{A}). When the result is extrapolated to $k \ell \lesssim 1$, this prefactor could vary and, in general, its dependence on $k \ell$ is likely to be more complex than just $1/(k \ell)^2$. In addition, the SC theory neglects interference processes insensitive to the breakdown of time-reversal invariance by, e.g., a strong magnetic field. The inclusion of such processes in the theoretical description would at least change the prefactor in Eq.\ (\ref{diffusion_coeff}). In Refs.\ \onlinecite{SB1,SB2,Nicolas}, for example, a larger prefactor was used in Eq.\ (\ref{diffusion_coeff}) to study the vicinity of the localization transition. This was justified by a comparison of some of the final results with those of the supersymmetric $\sigma$-model \cite{mirlin00}. Such a comparison indicates that the prefactor $6\pi/k^2 \ell$ in Eq.\ (\ref{diffusion_coeff}) have to be multiplied by 2 to obtain an exact correspondence between the two theoretical approaches  \cite{SB1} . 

Finally, SC theory of localization is a very convenient tool for description of realistic experimental situations, like the recent experiments on Anderson localization of light \cite{Maret}, microwaves \cite{zhang07}, ultrasound \cite{Page}, and matter waves \cite{lye05,clement05}. It can be adapted to almost any detail of a particular experiment (short pulses or focused beams, internal reflections on the sample surface, complex shapes or inhomogeneous scatterer density profiles of disordered samples, etc.). This gives SC theory a serious advantage as compared to other theories of Anderson localization.

{\it Note.} After this paper was submitted for publication, we became aware of the work of C. Tian \cite{tian08} who justifies the concept of the position-dependent diffusion coefficient using methods of supersymmetric field theory. 

\acknowledgments
We thank Bart van Tiggelen for many fruitful discussions.
SES acknowledges financial support from the French ANR (project 06-BLAN-0096 CAROL) and the French Ministry of Education and Research.

\appendix
\section{}
\label{A}

\begin{figure}[t]
\includegraphics[width=12cm]{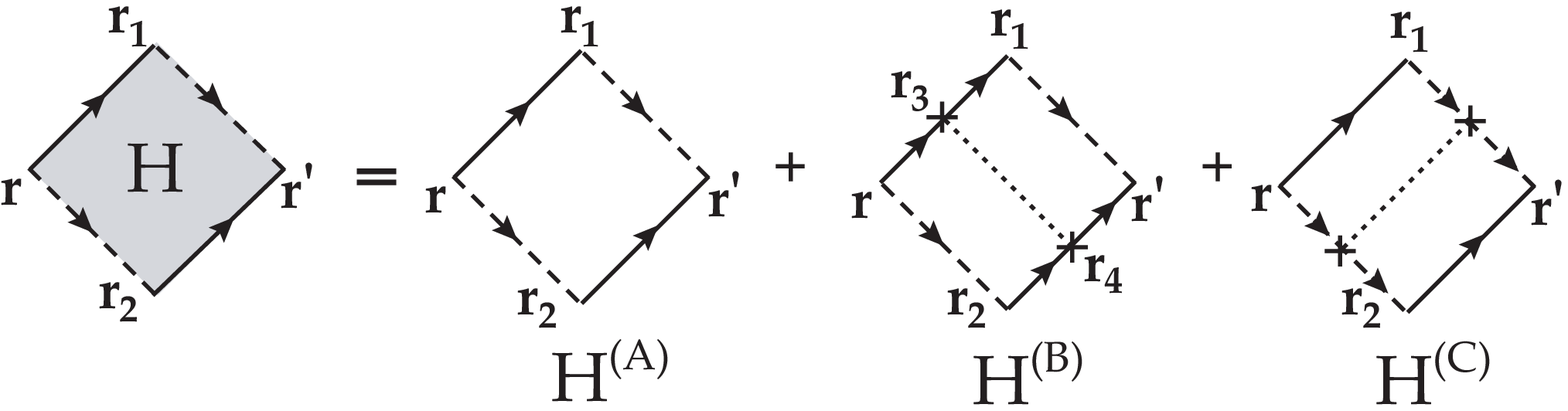}
\caption{Hikami box $H(\textbf{r},\textbf{r}_1,\textbf{r}^{\prime},\textbf{r}_2)$. Diagrammatic notation is the same as in Fig. \ref{Ladder_diagrams}.}
\label{square_boxes}  
\end{figure}

In this appendix we present a demonstration of Eq. (\ref{H_f_result}). Consider Eq. (\ref{H_f_definition}), where the Hikami box $H(\textbf{r},\textbf{r}_1,\textbf{r}^{\prime},\textbf{r}_2)$ is shown Fig.\ \ref{square_boxes}. This diagram is a sum of three contributions: $H^{(A)}(\textbf{r},\textbf{r}_1,\textbf{r}^{\prime},\textbf{r}_2)$, $H^{(B)}(\textbf{r},\textbf{r}_1,\textbf{r}^{\prime},\textbf{r}_2)$ and $H^{(C)}(\textbf{r},\textbf{r}_1,\textbf{r}^{\prime},\textbf{r}_2)$.
We hence have to perform three integrals. The second one, for example, is
\begin{equation}
\textbf{H}_f^{(B)}(\textbf{r},\textbf{r}^{\prime})=\int d\textbf{r}_1d\textbf{r}_2(\textbf{r}_1+\textbf{r}_2-2\textbf{r})H^{(B)}(\textbf{r},\textbf{r}_1,\textbf{r}^{\prime},\textbf{r}_2).
\label{The_integral}
\end{equation}
The two other integrals $\textbf{H}_f^{(A)}$ and $\textbf{H}_f^{(C)}$ are defined similarly.
In the following we focus on the calculation of $\textbf{H}_f^{(B)}$, the calculation being similar for $\textbf{H}_f^{(A)}$ and $\textbf{H}_f^{(C)}$. Eq. (\ref{The_integral}) can be rewritten as
\begin{eqnarray}
\textbf{H}_f^{(B)}(\textbf{r},\textbf{r}^{\prime}) &=& \dfrac{4\pi}{\ell}\int d\textbf{r}_1d\textbf{r}_2d\textbf{r}_3(\textbf{r}_1+\textbf{r}_2-2\textbf{r}) \langle G(\textbf{r},\textbf{r}_3)\rangle\langle G(\textbf{r}_3,\textbf{r}_1)\rangle \langle G^*(\textbf{r}_1,\textbf{r}^{\prime})\rangle
\nonumber \\
&\times&
 \langle G^*(\textbf{r},\textbf{r}_2)\rangle \langle G(\textbf{r}_2,\textbf{r}_3)\rangle \langle G(\textbf{r}_3,\textbf{r}^{\prime})\rangle.
\label{explicit_integral}
\end{eqnarray}
In this Appendix, we neglect the difference in frequencies $\omega_1$ and $\omega_2$ in the arguments of $\langle G \rangle$ and $\langle G^* \rangle$, respectively, and set $\omega_1 = \omega_2 = \omega_0$ for all amplitude Green's functions. This is justified as far as slow dynamics ($\Omega = \omega_1 - \omega_2 \ll \omega_0$, $c/\ell$) is concerned. To lighten the notation, we omit the frequency argument of $\langle G \rangle$.

By replacing the Green's functions in Eq. (\ref{explicit_integral}) by their Fourier transforms, we obtain
\begin{eqnarray}
\textbf{H}_f^{(B)}(\textbf{r},\textbf{r}^{\prime}) &=& \dfrac{4\pi}{\ell(2\pi)^{18}}\int d\textbf{r}_1d\textbf{r}_2d\textbf{r}_3d\textbf{k}_1\dots d\textbf{k}_6\langle G(\textbf{k}_1)\rangle\langle G(\textbf{k}_2)\rangle\langle G^*(\textbf{k}_3)\rangle
\nonumber \\
&\times&
\langle G^*(\textbf{k}_4)\rangle\langle G(\textbf{k}_5)\rangle\langle G(\textbf{k}_6)\rangle
 (\textbf{r}_1+\textbf{r}_2-2\textbf{r})
 \nonumber \\
 &\times& e^{-i\textbf{r}_3(\textbf{k}_1-\textbf{k}_2+\textbf{k}_5-\textbf{k}_6)}e^{-i\textbf{r}_1(\textbf{k}_2-\textbf{k}_3)}e^{-i\textbf{r}_2(\textbf{k}_4-\textbf{k}_5)}e^{-i\textbf{r}(-\textbf{k}_1-\textbf{k}_4)}e^{-i\textbf{r}^{\prime}(\textbf{k}_3+\textbf{k}_6)}\nonumber\\
&=& \mathbf{K}_1(\textbf{r},\textbf{r}^{\prime})+\mathbf{K}_2(\textbf{r},\textbf{r}^{\prime})+\mathbf{K}(\textbf{r},\textbf{r}^{\prime}),
\label{Hf_Fourier}
\end{eqnarray}
where $\mathbf{K}_1(\textbf{r},\textbf{r}^{\prime})$ is the part of Eq.\ (\ref{Hf_Fourier}) with the integrand proportional to $\textbf{r}_1$, $\mathbf{K}_2(\textbf{r},\textbf{r}^{\prime})$ is the part with the integrand proportional to $\textbf{r}_2$, and $\mathbf{K}(\textbf{r},\textbf{r}^{\prime})$ is the one with the integrand proportional to $-2\textbf{r}$. Let us first consider  $\mathbf{K}_1(\textbf{r},\textbf{r}^{\prime})$. In this term, the integrals over $\textbf{r}_2$ and $\textbf{r}_3$ give respectively $(2\pi)^3 \delta(\textbf{k}_4-\textbf{k}_5)$ and $(2\pi)^3 \delta(\textbf{k}_1-\textbf{k}_2+\textbf{k}_5-\textbf{k}_6)$, and the integral over $\textbf{r}_1$ gives $-i (2\pi)^3 \boldsymbol{\nabla}_{\textbf{k}_3}\delta(\textbf{k}_3-\textbf{k}_2)$. We have then
\begin{eqnarray}
\mathbf{K}_1(\textbf{r},\textbf{r}^{\prime}) &=& \dfrac{-4\pi i}{\ell(2\pi)^9}\int d\textbf{k}_1 d\textbf{k}_2 d\textbf{k}_4 \langle G(\textbf{k}_1)\rangle \langle G(\textbf{k}_2)\rangle \langle G^*(\textbf{k}_4)\rangle \langle G(\textbf{k}_4)\rangle \langle G(\textbf{k}_1-\textbf{k}_2+\textbf{k}_4)\rangle
\nonumber \\
&\times& e^{-i\textbf{r} (-\textbf{k}_1-\textbf{k}_4)}
\int d\textbf{k}_3 (\boldsymbol{\nabla}_{\textbf{k}_3} \delta(\textbf{k}_3-\textbf{k}_2))\langle G^*(\textbf{k}_3)\rangle e^{-i\textbf{r}^{\prime} (\textbf{k}_3+\textbf{k}_1-\textbf{k}_2+\textbf{k}_4)}.
\label{r1_term_1}
\end{eqnarray}
The integral over $\textbf{k}_3$ is equal to $-((\boldsymbol{\nabla}_{\textbf{k}_2}\langle G^*(\textbf{k}_2)\rangle)-i\textbf{r}^{\prime}\langle G^*(\textbf{k}_2)\rangle )e^{-i\textbf{r}^{\prime} (\textbf{k}_1+\textbf{k}_4)}$ and hence
\begin{eqnarray}
\mathbf{K}_1(\textbf{r},\textbf{r}^{\prime}) &=& \dfrac{4\pi i}{\ell(2\pi)^9}\int d\textbf{k}_1 d\textbf{k}_2 d\textbf{k}_4 \langle G(\textbf{k}_1)\rangle \langle G(\textbf{k}_2)\rangle \langle G^*(\textbf{k}_4)\rangle 
\nonumber \\
&\times&
\langle G(\textbf{k}_4)\rangle (\boldsymbol{\nabla}_{\textbf{k}_2}\langle G^*(\textbf{k}_2)\rangle) \langle G(\textbf{k}_1-\textbf{k}_2+\textbf{k}_4)\rangle e^{i(\textbf{r}-\textbf{r}^{\prime})(\textbf{k}_1+\textbf{k}_4)}\nonumber\\
&+& \dfrac{4\pi\textbf{r}^{\prime}}{\ell(2\pi)^9}\int d\textbf{k}_1 d\textbf{k}_2 d\textbf{k}_4 \langle G(\textbf{k}_1)\rangle \langle G(\textbf{k}_2)\rangle \langle G^*(\textbf{k}_4)\rangle
\nonumber \\
&\times&
 \langle G(\textbf{k}_4)\rangle \langle G^*(\textbf{k}_2)\rangle \langle G(\textbf{k}_1-\textbf{k}_2+\textbf{k}_4)\rangle e^{i(\textbf{r}-\textbf{r}^{\prime}) (\textbf{k}_1+\textbf{k}_4)}.
\label{r1_term_2}
\end{eqnarray}
The second term on the r.h.s. is nothing else than $\textbf{r}^{\prime}H^{(B)}(\textbf{r},\textbf{r}^{\prime})$, where
\begin{equation}
H^{(B)}(\textbf{r},\textbf{r}^{\prime})=\int d\textbf{r}_1 d\textbf{r}_2 H^{(B)}(\textbf{r},\textbf{r}_1,\textbf{r}^{\prime},\textbf{r}_2).
\end{equation}
In the first term on the r.h.s. of Eq. (\ref{r1_term_2}) we change the variables $\textbf{k}_1\rightarrow \textbf{k}$, $\textbf{k}_2\rightarrow \textbf{k}^{\prime}$, and $\textbf{k}_1+\textbf{k}_4\rightarrow \textbf{q}$. Equation (\ref{r1_term_2}) becomes
\begin{eqnarray}
\mathbf{K}_1(\textbf{r},\textbf{r}^{\prime}) &=& \dfrac{4\pi i}{\ell(2\pi)^9}\int d\textbf{k} d\textbf{k}^{\prime} d\textbf{q} \langle G(\textbf{k})\rangle \langle G(\textbf{k}^{\prime})\rangle \langle G^*(\textbf{q}-\textbf{k})\rangle
\nonumber \\
&\times&
 \langle G(\textbf{q}-\textbf{k})\rangle (\boldsymbol{\nabla}_{\textbf{k}^{\prime}}\langle G^*(\textbf{k}^{\prime})\rangle) \langle G(\textbf{q}-\textbf{k}^{\prime})\rangle e^{i\textbf{q} (\textbf{r}-\textbf{r}^{\prime})}
+ \textbf{r}^{\prime}H^{(B)}(\textbf{r},\textbf{r}^{\prime}).
\label{r1_term_3}
\end{eqnarray}
In the limit of small $\textbf{q}$, we have $1/(2\pi)^3 \int d\textbf{k}\langle G(\textbf{k})\rangle \langle G^*(\textbf{q}-\textbf{k})\rangle \langle G(\textbf{q}-\textbf{k})\rangle=-i\ell^2(1-q^2\ell^2/3)/8\pi k$ \cite{Akkermans_Montambaux} and 
\begin{eqnarray}
\mathbf{K}_1(\textbf{r},\textbf{r}^{\prime}) &=& \dfrac{1}{(2\pi)^3}\int d\textbf{q} e^{i\textbf{q} (\textbf{r}-\textbf{r}^{\prime})}\dfrac{\ell}{2k}\left( 1-\dfrac{q^2\ell^2}{3} \right)
\nonumber \\
&\times&
 \dfrac{1}{(2\pi)^3}\int d\textbf{k}^{\prime}\langle G(\textbf{k}^{\prime})\rangle \langle G(\textbf{q}-\textbf{k}^{\prime})\rangle (\boldsymbol{\nabla}_{\textbf{k}^{\prime}}\langle G^*(\textbf{k}^{\prime})\rangle)+\textbf{r}^{\prime}H^{(B)}(\textbf{r},\textbf{r}^{\prime}).
\label{K1}
\end{eqnarray}
A similar calculation gives
\begin{eqnarray}
\mathbf{K}_2(\textbf{r},\textbf{r}^{\prime}) &=& -\dfrac{1}{(2\pi)^3}\int d\textbf{q} e^{i\textbf{q} (\textbf{r}-\textbf{r}^{\prime})}\dfrac{\ell}{2 k}\left( 1-\dfrac{q^2\ell^2}{3} \right)
\nonumber \\
&\times&
\dfrac{1}{(2\pi)^3}\int d\textbf{k}^{\prime}\langle G(\textbf{k}^{\prime})\rangle \langle G(\textbf{q}-\textbf{k}^{\prime})\rangle (\boldsymbol{\nabla}_{\textbf{k}^{\prime}}\langle G^*(\textbf{k}^{\prime})\rangle)+\textbf{r}H^{(B)}(\textbf{r},\textbf{r}^{\prime}).
\label{K2}
\end{eqnarray}
Besides, it follows straightforwardly from Eq. (\ref{explicit_integral}) that $\mathbf{K}(\textbf{r},\textbf{r}^{\prime})=-2\textbf{r}H^{(B)}(\textbf{r},\textbf{r}^{\prime})$. Combined with Eqs. (\ref{K1}) and (\ref{K2}), this yields
\begin{eqnarray}
\textbf{H}^{(B)}_f(\textbf{r},\textbf{r}^{\prime}) &=& \mathbf{K}_1(\textbf{r},\textbf{r}^{\prime})+\mathbf{K}_2(\textbf{r},\textbf{r}^{\prime})+\mathbf{K}(\textbf{r},\textbf{r}^{\prime})
\nonumber \\
&=& -(\textbf{r}-\textbf{r}^{\prime})H^{(B)}(\textbf{r},\textbf{r}^{\prime}).
\label{The_integral_B}
\end{eqnarray}
The calculation of $\textbf{H}^{(A)}$ and $\textbf{H}^{(C)}$ follows the same lines. We obtain
\begin{eqnarray}
\textbf{H}_f^{(A)}(\textbf{r},\textbf{r}^{\prime}) &=& 
\int d\textbf{r}_1d\textbf{r}_2(\textbf{r}_1+\textbf{r}_2-2\textbf{r})H^{(A)}(\textbf{r},\textbf{r}_1,\textbf{r}^{\prime},\textbf{r}_2)
\nonumber \\
&=& -(\textbf{r}-\textbf{r}^{\prime})H^{(A)}(\textbf{r},\textbf{r}^{\prime})
\label{The_integral_A}
\end{eqnarray}
and
\begin{eqnarray}
\textbf{H}_f^{(C)}(\textbf{r},\textbf{r}^{\prime}) &=& \int d\textbf{r}_1d\textbf{r}_2(\textbf{r}_1+\textbf{r}_2-2\textbf{r})H^{(C)}(\textbf{r},\textbf{r}_1,\textbf{r}^{\prime},\textbf{r}_2)
\nonumber \\
&=& -(\textbf{r}-\textbf{r}^{\prime})H^{(C)}(\textbf{r},\textbf{r}^{\prime}).
\label{The_integral_C}
\end{eqnarray}
Combining Eqs.\ (\ref{The_integral_B}), (\ref{The_integral_A}), and (\ref{The_integral_C}) we find
\begin{eqnarray}
\textbf{H}_f(\textbf{r},\textbf{r}^{\prime}) &=& \textbf{H}_f^{(A)}(\textbf{r},\textbf{r}^{\prime})+\textbf{H}_f^{(B)}(\textbf{r},\textbf{r}^{\prime})+\textbf{H}_f^{(C)}(\textbf{r},\textbf{r}^{\prime})
\nonumber \\
&=& -(\textbf{r}-\textbf{r}^{\prime})H(\textbf{r},\textbf{r}^{\prime}),
\end{eqnarray}
which is Eq. (\ref{H_f_result}) of the main text.

\section{}
\label{B}
We show here that the inverse Fourier transform of Eq. (\ref{Cloop_r_Dr_Fourier}) with respect to $\textbf{q}$ is given by Eq. (\ref{Cloop_r_Dr_explicit}).
As a function of $\mathbf{q}$, Eq.\ (\ref{Cloop_r_Dr_Fourier}) is a sum of two terms proportional to $(i \mathbf{q})^2$ and $i \mathbf{q}$, respectively. The inverse Fourier transform of $i \mathbf{q}$ is
\begin{eqnarray}
\int\frac{d \mathbf{q}}{(2 \pi)^3}\; i \mathbf{q}\; e^{i \mathbf{q} \cdot \Delta \mathbf{r}} =
\boldsymbol{\nabla}_{\Delta \mathbf{r}} \left[
\int\frac{d \mathbf{q}}{(2 \pi)^3}\; e^{i \mathbf{q} \cdot \Delta \mathbf{r}}
\right]
= \boldsymbol{\nabla}_{\Delta \mathbf{r}} \delta(\Delta \mathbf{r}).
\label{ftiq}
\end{eqnarray}
Similarly, the inverse Fourier transform of $(i \mathbf{q})^2$ is
$\Delta_{\Delta \mathbf{r}} \delta(\Delta \mathbf{r})$. This leads directly to Eq.\ (\ref{Cloop_r_Dr_explicit}).

\section{}
\label{C}
Here we obtain Eq. (\ref{applied_operator_init}) from the series of Eq. (\ref{Resummation}). The idea is to apply the operator $-i\Omega-D_B\Delta_{\textbf{r}}$ to both sides of Eq. (\ref{Resummation}). The first term on the r.h.s. is transformed into $\delta(\textbf{r}-\textbf{r}^{\prime})$ since $C_D(\textbf{r},\textbf{r}_1,\Omega)$ obeys Eq. (\ref{eq_for_C_infinite}), and for each of the next terms the first multiplier $C_D(\textbf{r},\textbf{r}_1,\Omega)$ in the integrands is transformed into $\delta(\textbf{r}-\textbf{r}_1)$ for the same reason. Equation (\ref{Resummation}) becomes
\begin{eqnarray}
\left[-i\Omega-D_B\Delta_{\textbf{r}} \right]C(\textbf{r},\textbf{r}^{\prime},\Omega) &=&  \delta(\textbf{r}-\textbf{r}^{\prime})+ \frac{4 \pi c}{\ell^2} \int \delta(\textbf{r}-\textbf{r}_1)X(\textbf{r}_1,\textbf{r}_2,\Omega)C_D(\textbf{r}_2,\textbf{r}^{\prime},\Omega)d\textbf{r}_1 d\textbf{r}_2\nonumber\\
&+& \left( \frac{4 \pi c}{\ell^2} \right)^2 \int \delta(\textbf{r}-\textbf{r}_1)X(\textbf{r}_1,\textbf{r}_2,\Omega)C_D(\textbf{r}_2,\textbf{r}_3,\Omega)
\nonumber \\
&\times&
X(\textbf{r}_3,\textbf{r}_4,\Omega)C_D(\textbf{r}_4,\textbf{r}^{\prime},\Omega)d\textbf{r}_1 d\textbf{r}_2d\textbf{r}_3d\textbf{r}_4 + \dots
\end{eqnarray}
Performing integrations over $\textbf{r}_1$ we obtain
\begin{eqnarray}
\left[-i\Omega-D_B \Delta_{\textbf{r}} \right]C(\textbf{r},\textbf{r}^{\prime},\Omega) &=& \delta(\textbf{r}-\textbf{r}^{\prime})+ \dfrac{4 \pi c}{\ell^2} \int X(\textbf{r},\textbf{r}_2,\Omega)C_D(\textbf{r}_2,\textbf{r}^{\prime},\Omega)d\textbf{r}_2\nonumber\\
&+& \left(\dfrac{4 \pi c}{\ell^2} \right)^2\int X(\textbf{r},\textbf{r}_2,\Omega)C_D(\textbf{r}_2,\textbf{r}_3,\Omega)
\nonumber \\
&\times&
X(\textbf{r}_3,\textbf{r}_4,\Omega)C_D(\textbf{r}_4,\textbf{r}^{\prime},\Omega)d\textbf{r}_2d\textbf{r}_3d\textbf{r}_4 + \dots
\label{serie_sans_delta}
\end{eqnarray}
Now let us perform integrations over $\textbf{r}_2$. We have to calculate an integral
\begin{equation}
I = \dfrac{4 \pi c}{\ell^2} \int X(\textbf{r},\textbf{r}_2,\Omega)C_D(\textbf{r}_2,\textbf{r}_3,\Omega)d\textbf{r}_2,
\end{equation}
where $\textbf{r}_3=\textbf{r}^{\prime}$ for the first integral on the r.h.s. of Eq.\ (\ref{serie_sans_delta}). To this end, we use Eq. (\ref{Cloop_r_r'_explicit}) for $X(\textbf{r},\textbf{r}_2,\Omega)$ and obtain
\begin{eqnarray}
I &=& \dfrac{-\ell^2D_B}{2k^2}\int d\textbf{r}_2 \left[ \boldsymbol{\nabla}_{\textbf{r}} \cdot \Gamma_D(\textbf{r},\textbf{r},\Omega)\boldsymbol{\nabla}_{\textbf{r}} \delta(\textbf{r}-\textbf{r}_2) \right] C_D(\textbf{r}_2,\textbf{r}_3,\Omega) \nonumber\\
&=& \dfrac{-\ell^2D_B}{2k^2}\boldsymbol{\nabla}_{\textbf{r}} \cdot \left[\Gamma_D(\textbf{r},\textbf{r},\Omega)\boldsymbol{\nabla}_{\textbf{r}}\right]C_D(\textbf{r},\textbf{r}_3,\Omega),
\end{eqnarray}
where we integrated by parts.

Replacing each integration over $\textbf{r}_2$ by this result in Eq. (\ref{serie_sans_delta}) we obtain
\begin{eqnarray}
\left[-i\Omega-D_B \Delta_{\textbf{r}} \right]C(\textbf{r},\textbf{r}^{\prime},\Omega) &=&
\delta(\textbf{r}-\textbf{r}^{\prime})
\nonumber \\
&-& \dfrac{\ell^2D_B}{2k^2}\boldsymbol{\nabla}_{\textbf{r}} \cdot \Gamma_D(\textbf{r},\textbf{r},\Omega)\boldsymbol{\nabla}_{\textbf{r}}
\Big[ C_D(\textbf{r},\textbf{r}^{\prime},\Omega)
\nonumber \\
&+& \frac{4 \pi c}{\ell^2} \int C_D(\textbf{r},\textbf{r}_3,\Omega)X(\textbf{r}_3,\textbf{r}_4,\Omega)C_D(\textbf{r}_4,\textbf{r}^{\prime},\Omega)d\textbf{r}_3 d\textbf{r}_4\nonumber\\
&+& \left( \frac{4 \pi c}{\ell^2} \right)^2 \int C_D(\textbf{r},\textbf{r}_3,\Omega)X(\textbf{r}_3,\textbf{r}_4,\Omega)C_D(\textbf{r}_4,\textbf{r}_5,\Omega)
\nonumber \\
&\times&
X(\textbf{r}_5,\textbf{r}_6,\Omega)C_D(\textbf{r}_6,\textbf{r}^{\prime},\Omega)d\textbf{r}_3d\textbf{r}_4d\textbf{r}_5d\textbf{r}_6
+ \dots \Big]
\label{last_series}
\end{eqnarray}
The infinite series in square brackets is nothing else than the intensity Green's function $C(\textbf{r},\textbf{r}^{\prime},\Omega)$ as given by Eq. (\ref{Resummation}). Thus, Eq. (\ref{last_series}) leads straightforwardly to Eq. (\ref{applied_operator_init}).

\section{}
\label{D}
We prove here that the proportionality between $C_D$ and $\Gamma_D$ known in the infinite medium \cite{Akkermans_Montambaux} holds in a finite medium as well. Calculations being similar to those of Appendix \ref{A}, we only give the main ingredients of the proof. 
According to Eq. (\ref{Rigorous_intensity}), at $|\mathbf{r} - \mathbf{r}^{\prime}| \gg \ell$, $C_D(\textbf{r},\textbf{r}^{\prime},\Omega)$ is given by
\begin{equation}
C_D(\textbf{r},\textbf{r}^{\prime},\Omega)=\dfrac{4\pi}{c}\int d\textbf{r}_1d\textbf{r}_2\langle G(\textbf{r},\textbf{r}_1)\rangle\langle G^*(\textbf{r},\textbf{r}_1)\rangle\Gamma_D(\textbf{r}_1,\textbf{r}_2,\Omega)\langle G(\textbf{r}_2,\textbf{r}^{\prime})\rangle\langle G^*(\textbf{r}_2,\textbf{r}^{\prime})\rangle.
\label{C_D_definition}
\end{equation}
Similarly to Appendix \ref{A}, we omit frequency arguments of amplitude Green's functions and set all of them equal to $\omega_0$.
Because $\langle G(\mathbf{r}, \mathbf{r}_1) \rangle$ is exponentially small for $|\mathbf{r} - \mathbf{r}_1| > \ell$, the main contribution to the integral comes from $|\mathbf{r} - \mathbf{r}_1|$, $|\mathbf{r}^{\prime} - \mathbf{r}_2| < \ell$. This authorizes us to expand $\Gamma_D(\textbf{r}_1,\textbf{r}_2,\Omega)$ in series around $(\textbf{r},\textbf{r}^{\prime})$. This expansion has to be truncated to the same first order in $|\mathbf{r} - \mathbf{r}_1|$ and $|\mathbf{r}^{\prime} - \mathbf{r}_2|$ as the expansion of Eq. (\ref{expansion_simplified}):
\begin{equation}
\Gamma_D(\textbf{r}_1,\textbf{r}_2,\Omega)\simeq \Gamma_D(\textbf{r},\textbf{r}^{\prime},\Omega)+(\textbf{r}_1+\textbf{r}_2-\textbf{r}-\textbf{r}^{\prime})
\cdot \boldsymbol{\nabla}_{\textbf{r}}\Gamma_D(\textbf{r},\textbf{r}^{\prime},\Omega).
\label{expansion2}
\end{equation} 
We then substitute Eq.\ (\ref{expansion2}) into Eq.\ (\ref{C_D_definition}). The integral proportional to $\Gamma_D(\textbf{r},\textbf{r}^{\prime},\Omega)$ is the usual result obtained in the infinite medium. This integral equals $\ell^2/(4\pi c) \Gamma_D(\textbf{r}_1,\textbf{r}_2,\Omega)$ \cite{Akkermans_Montambaux}. We hence obtain
\begin{eqnarray}
\label{C_D_2}
C_D(\textbf{r},\textbf{r}^{\prime},\Omega) &=& \dfrac{\ell^2}{4\pi c}\Gamma_D(\textbf{r},\textbf{r}^{\prime},\Omega)
\\
&+& \dfrac{4\pi}{c} \left[ \int d\textbf{r}_1d\textbf{r}_2(\textbf{r}_1+\textbf{r}_2-\textbf{r}-\textbf{r}^{\prime}) \left| \langle G(\textbf{r},\textbf{r}_1)\rangle \right|^2 \left| \langle G(\textbf{r}_2,\textbf{r}^{\prime})\rangle \right|^2 \right] \cdot \boldsymbol{\nabla}_{\mathbf{r}}\Gamma_D(\textbf{r},\textbf{r}^{\prime}, \Omega). \nonumber
\end{eqnarray}
The integral on the r.h.s. of Eq. (\ref{C_D_2}) can be calculated exactly in the same way as $\textbf{H}_f^{A}$, $\textbf{H}_f^{B}$ or $\textbf{H}_f^{C}$ in Appendix \ref{A}, and it is easy to see that this integral is zero.

\end{document}